\documentclass[journal]{IEEEtran}
\IEEEoverridecommandlockouts
\usepackage{cite}
\usepackage{amsmath,amssymb,amsfonts}
\usepackage{booktabs, makecell, tabularx}
\usepackage{algorithmic}
\usepackage{graphicx}
\usepackage{textcomp}
\usepackage{xcolor}
\def\BibTeX{{\rm B\kern-.05em{\sc i\kern-.025em b}\kern-.08em
    T\kern-.1667em\lower.7ex\hbox{E}\kern-.125emX}}
\usepackage{caption} 
\usepackage[utf8]{inputenc}

\usepackage[english]{babel}
\usepackage[letterpaper, margin=0.57in, tmargin=0.7in]{geometry}
\begin{document}

\title{A Survey on IoT Smart Healthcare: Emerging Technologies, Applications, Challenges, and Future Trends  }

\author{\IEEEauthorblockN{M. Ali Tunç\IEEEauthorrefmark{1},
Emre Gures\IEEEauthorrefmark{2}, Ibraheem Shayea\IEEEauthorrefmark{3}}

\IEEEauthorblockA{Department of Electrical and Electronics Engineering,
Istanbul Technical University\\
Istanbul, Turkey\\
Email: \IEEEauthorrefmark{1}tuncmuh@itu.edu.tr,
\IEEEauthorrefmark{2}gurese18@itu.edu.tr,
\IEEEauthorrefmark{3}ibr.shayea@gmail.com}}
\maketitle

\begin{abstract}
 The internet of things (IoT) refers to a framework of interrelated, web associated objects that can gather and move information over a remote network without human interference. With a quick development in the arrangement of IoT gadgets and expanding want to make medical care more financially savvy, customized, and proactive, IoT is ready to assume a solid function in all perspectives of the healthcare industry. In this context, IoT-based healthcare provides several benefits such as instant and reliable treatment, cost reduction, improved communication, etc. by using different new technologies. Wireless Body Area Networks (WBAN) technologies can enhance the quality of data gathering and data transferring in smart systems. Machine Learning(ML) are put to use at every level of smart healthcare systems. Fog computing reduces communication cost and provides low latency. Software-Defined Networking (SDN) and Network Function Virtualization (NFV) technologies provide less complex and more flexible network structures. Blockchain technology offers a better way of protection of users' sensitive information. This paper aims to provide comprehensive descriptions of ongoing research projects and the utilization of the above-mentioned technologies in smart healthcare systems. In this paper, the latest studies, proposed techniques, and the current solutions of smart healthcare systems are elaborated in the context of emerging technologies, applications and challenges of these systems to provide a better understanding of what IoT means in the healthcare industry now and what it will mean in the future. \footnote{This research has been produced benefiting from the 2232 International Fellowship for Outstanding Researchers Program of TÜBİTAK (Project No: 118C276) conducted at Istanbul Technical University (ITU).}
\end{abstract}.


\section{Introduction}

\IEEEPARstart{W}{ith} the rapid growth of communication and computer technologies, the IoT term is taking up more and more space in our daily life day by day and it is estimated that by 2020, 25 billion "things" will be connected to the internet \cite{1}. In recent years, the healthcare industry realized that there has been an expanding interest in creating effective and practical medical care arrangements which encourage medical clinics and medical care suppliers ease in performing finding and encouraging patients. And hence; to obtain solutions to smart healthcare challenges, the average spending on smart healthcare in Europe is approximately 10\% of gross domestic product (GDP), and up to 99 billion euros of healthcare costs can be saved through smart healthcare by 2020 \cite{3}. IoT is a megatrend of cutting edge innovations affecting the entire business range and is an interconnection of exceptionally recognizable shrewd items and gadgets.

The developments of IoT and cloud technology has empowered pervasive communication by giving availability across different objects for sharing information. IoT offers a stage to associate heterogeneous gadgets and gives preparing and cooperation capacities. It is conveyed in a few zones to assemble smart homes, smart urban communities, smart healthcare, and so on. IoT empowers imagining objects that can detect, measure, impart and share data over the organization of gadgets. These interconnected objects and sensors gather the information which is dissected and utilized for different purposes like classification, decision making, arranging and so forth\cite{69}.

The healthcare sector is demanding innovations and applications to provide proper and low latency healthcare and also new applications are mandatory for handling the rising number of patients. In many ways, traditional healthcare has issues to overcome for a sufficient and reliable healthcare service at the individual and clinical level. At first, traditional healthcare services have difficulties in ensuring the quality of services demands of individuals as the growth of the population rises. And also, in traditional healthcare systems, the lack of medical staff causes late diagnosis and inadequate treatment process for the patients. In addition to that, treatment of some of the critical and insidious diseases such as cancer, Alzheimer's disease, etc., needs to early detection process of the diseases and traditional healthcare services may fail in this task because of the insufficient and deficient examination of diseases. Besides, the emergency response of some illnesses may be difficult to be ensured by traditional healthcare services and this may lead to deadly consequences. Lastly, it can be said that in rural areas, transportation to a centralized hospital from small towns may take a long duration which can cause undesired results for critical patients.

With the emerging wireless communication technologies, sensor technologies and above all IoT technologies based on 5G, the home care and remote healthcare applications for the low, medium and high-risk patients can be applied easily. In this context, smart healthcare term has turned up into people's daily lives for easing and enhancing people's health care methods.

 Smart healthcare is a healthcare administration framework that utilizes innovations such as wearable gadgets, IoT, and different network types to powerfully get to data, interface individuals, materials, and foundations identified with medical care, and afterwards, effectively oversee and react to the clinical environment needs intelligently. Smart healthcare systems can use various technologies such as fog computing, cloud computing, ML, SDN, artificial intelligence (AI), blockchain technology, etc.

 With emerging technologies such as cognitive computing, artificial intelligence, cloud computing, fog computing, edge computing, so forth; smart healthcare systems are considered to be pioneer solutions to all challenges that the healthcare industry suffers. Smart healthcare systems can be used for analysis and summarization of the huge amount of data and correlation between the analysis of patients' medical data and the other factors such as age, gender, the environment of patients. For example; in \cite{52}, it is introduced that remote healthcare monitoring systems (RHMS) which are all in all three levels of design where the first level uses astute wearable sensors to accumulate physiological signs. Most wearable sensors constructors marketed detecting gadgets with Bluetooth low energy (BLE) correspondence interfaces, which lead to the advancement of different remote healthcare monitoring systems sending BLE correspondence interfaces for physiological patient information gathering. In many cases, due to their basic status, now and then chronic patients' well-being goes unnoticed until the infections form into an emergency stage and smart healthcare systems can provide instant diagnosis and rapid treatment. In \cite{51}, it is proposed that a self-adaptative early admonition scores framework that regards a hazard assessment approach. It gives a manual and self-adaptative setup of the crucial signs observing cycle contingent upon the patient well-being status variety and the clinical staff choices. Some chronic diseases and critical conditions of the patients are being monitored at their homes and it may be needed to make decisions quickly and give emergency alerts at instant changes in the patients' health status. The authors proposed a novel system for coordinating outfit profound learning in Edge registering gadgets and conveyed it for genuine use of programmed Heart Disease examination\cite{54}. This novel system conveys medical care as a haze administration utilizing IoT gadgets and proficiently deals with the information of heart patients, which comes as client demands. By employing AI and machine learning algorithms, IoT has explicitly made it achievable to anticipate the beginning of deadly episodes. For example, an armband with the sensors for the discovery of the surface electromyography signal (sEMG) is planned in \cite{49} and the information gathered by these factors is handled by ML-based classification estimating algorithms and principal component analysis. This framework can distinguish the motions from the sEMG signals with a precision of around 97\%.
 
Along with all of the benefits of smart healthcare systems, there are several numbers of challenges to be solved and enhanced eventually. The generation of the huge volume of data by sensors and IoT gadgets is essential for providing a better analysis of the patients' health status. This huge amount of data can not easily be processed and also produce a high energy consumption. Low-latency outcomes and diagnosis are considered life-savers for high-risk patients especially. On the contrary of the developers and manufacturers of smart healthcare systems, the users of these systems which are doctors, patients, caregivers can't have an obligation for understanding the details of smart healthcare systems so systems' outcomes and alert mechanisms should be user-friendly. One of the main challenges of smart healthcare is the difficulty of ensuring the privacy and security of these systems. the hackers and the third parties are intrigued by the medical data of the users which consists of social security numbers, health status, the user's privacy records, etc. Unless the solutions to each above-mentioned challenges are occurred and deployed to the real world's environments, It shouldn't be said that smart healthcare is a new future for the healthcare industry.
 
To fully understanding smart healthcare; first, we must understand the different elements and aspects of smart healthcare. In this paper, first, we presented that what IoT and smart healthcare are in section I and then discussed the emerging technologies in smart healthcare in section II. The applications and services were presented in section III. Then, the challenges of smart healthcare and the latest proposals and techniques that were offered to solve these challenges were examined in section IV. Finally, future trends in smart healthcare were addressed in section V and we discussed our conclusions in section VI. And also, a literature comparison between this paper and forefront survey papers in recent years are described in Table I.

\section{IoT Smart Healthcare}

IoT is the network that contains a variety of physical objects with embedded technologies for communication and sensing process.
IoT term offers improvements to the modern people for easing their lives by IoT devices such as sensors, actuators, smartphones, etc. The growth of mobile users, the development in communication technologies and the raising capacities of cloud technologies bring the smart city, smart home, smart healthcare concepts into people' daily lives. IoT provides a global ecosystem where IoT devices can share their measured data and with cloud servers to accomplish their determined goals without human interaction and cooperate to create new applications. Different protocols and communication types such as Bluetooth, ZigBee, Wifi, 5G and so on, can be utilized in IoT networks according to network requirements, user satisfaction and wireless communication ranges. The generated and transferred data can be analyzed and used for classification, decision making and planning, etc. According to the purpose of the utilization of the measured data, the outcomes of the data, opening a door, changing the room temperature, alerting abnormal situations, changing traffic lights, and so on, are evaluated instantly or, for example, the generation of databases, creating useful results from the raw data, are transmitted to cloud servers or third party components of the network. 

The development of 5G wireless communication systems, which has paved the way for providing high throughput and low latency for their users, can be considered as a key factor in the enhancement of IoT applications and services.

\begin{table*}[!t]
   \caption{Literature survey comparison.}

   \begin{tabularx}{\textwidth}{p{0.9cm} p{0.4cm} p{15cm}}
     \toprule
      \ Survey &Year  & Description \\

 \midrule
 
 \cite{95} & 2017 &  • The latest IoT healthcare applications, platforms are discussed and the emerging technologies which are utilized for networking, data processing and sensing in IoT healthcare systems are presented in this paper. 
 
  • The fundamental problems related to the high volume of data, sensing interoperability, intelligent data processing mechanisms, uncontrolled environments etc., are presented in the context of technical challenges and future trends of IoT healthcare systems.
 \hspace{1sp}
 \\
 \midrule
 \cite{93} &2018 & • The requirements of IoT-based healthcare applications for a reliable end-to-end communication process are introduced and the emerging communication technologies which can fulfil these requirements are presented.
 
 • The survey is conducted in four different application scenarios based on different diseases such as infectious diseases, cardiovascular diseases, musculoskeletal disorders and neuromuscular disorders, respectively.
 
 • Open issues and challenges of IoT-based healthcare applications are mentioned and the fulfilment of requirements in different healthcare applications are discussed by the exploration of the emerging communication technologies and standards.
 
   \hspace{1sp}
 \\
 \midrule
 \cite{98} & 2018 & • The various IoT-based healthcare systems are discussed at a technical and practical level in the view of four main topics such as e-health systems, telemedicine and home monitoring systems, RFID based monitoring systems and Named Data Networking (NDN) based healthcare systems.
 
 • It is presented that these above-mentioned systems can be utilized for providing sufficient and cost-effective healthcare service at the individual level and overcoming the issues of smart healthcare systems such as reliability, low-latency tolerance, interoperability and many more.

    \hspace{1sp}
 \\
 \midrule

 \cite{97} & 2018 & • The main application areas of IoT-based healthcare systems and the fundamental communication technologies and protocols are presented briefly.
 
 • The overall outline of this paper is based on predetermined questions in the view of the essential applications and network elements, the most important technologies, the main issues and challenges of IoT healthcare systems, respectively.
 
 • The main components and their functions in the latest proposals of IoT-based healthcare systems are compared in a table at the end of this paper.
 \hspace{1sp}
 \\
 \midrule
  \cite{92} & 2019 & • An extended review of IoT-based applications and services in smart healthcare systems is presented. Different applications and services are described in the view of their main aims and fields of usage to offer a useful insight into smart healthcare systems.

    \hspace{1sp}
 \\
 \midrule
 
 \cite{91} & 2019 & • An overall review of the state-of-art in the context of different architecture components such as sensing, communication and data analytics of the current IoT healthcare systems is submitted in this paper.
 
 • The challenges which IoT healthcare systems face are reviewed and open issues, future trends and emerging directions are discussed.
    \hspace{1sp}
 \\
 \midrule
 \cite{94} & 2019 &   • The research is conducted based on four topics such as healthcare service delivery,  pharmaceutical industry, health monitoring and e-health, respectively. 5 papers for each topic are selected from the literature research and are mentioned with their advantages and drawbacks.
 
 •The advantages and weaknesses of other related surveys are presented and the literature research mechanism for this paper is presented in detail. 
 
 • At the end of this paper, the authors indicate their reviews and the open issues in IoT healthcare systems for future works.  
      \hspace{1sp}
 \\
 \midrule
 
 \cite{10} & 2019 & • IoT applications and WBAN-based IoT technologies which can ensure the requirements of IoT healthcare systems are presented. And also, the network architecture of WBAN-based systems are explained briefly.
 
 • The standards, specifications, advantages and disadvantages of the latest WBAN-based healthcare applications are mentioned in a table briefly.
 
 • After mentioning IoT healthcare services and applications, open issues the main challenges are discussed at the end of this paper.
 
   \hspace{1sp}
 \\
 \midrule
 
 \cite{99} & 2020 & • An overall view of the architecture of smart healthcare systems and the main requirements of these systems are presented.
 
 • The outline of this paper is based on various emerging technologies which are used to enhance the performances of smart healthcare systems such as Big Data, ML, blockchain and SDN. Many related studies of the utilization of these technologies in smart healthcare are discussed under different topics.
 
 • Future trends and open issues are reviewed at the end of this paper to provide making sufficient and rapid improvements in smart healthcare systems.
  \hspace{1sp}
 \\
 \midrule
 This work & 2021 & • The requirements of traditional healthcare systems, an overall concept of the smart healthcare network infrastructure and the current conditions of the emerging technologies which are used in smart healthcare systems are introduced.

• The main applications and services, and the challenges of smart healthcare are discussed to give a proper understanding of the requirements and the functions of smart healthcare systems. 

• At the end of the paper, the future trends of smart healthcare are presented and reviewed in the context of the mentioned technologies such as fog computing, ML, SDN, NFV and blockchain to navigate the researchers for enhancing healthcare systems.
\\

 \bottomrule

\end{tabularx}
\end{table*}

Due to the increasing number of connected devices and the massive growth of data volume generated by IoT sensors, IoT without 5G technologies cannot meet the user's demands for high Quality of Service (QoS). At this point, 5G can support IoT systems in the context of increasing throughput, transmission coverage, energy efficiency, reliability, and reducing delay.
With the interaction of different kinds of smart devices, IoT brings adaptability and comfort in conveying in different conditions for observing and communication purposes \cite{69}. Embedded sensors and IoT devices that measure ECG signals, blood pressure, body temperature, oxygen saturation level, body movements, etc. can be utilized on a patient's body or in hospital and home environments. Smart healthcare enables various information which can be related to an individual's health status to utilize for diagnosing diseases. The utilization of the smart healthcare frameworks with various IoT abilities allows distant monitoring and continuous following of patient's medical issue, long term review of patient's wellbeing records, decreasing clinical costs and expanding the innovation for giving patient-driven care rather than medical clinic-driven treatment.

Smart Healthcare infrastructure can be divided into five units such as physiological sensor unit, processing unit, communication and transmission unit, storage and computing unit, data analytics and decision making unit \cite{84}. Smart healthcare devices not only measure and monitor the physiological data but also process and transmits the medical data to remote healthcare or other IoT devices. Smart healthcare devices communicate through short-range communication technologies such as Bluetooth, Zigbee, Wifi, etc. to other IoT devices or through long-range communication technologies such as Worldwide Interoperability for Microwave Access (WiMAX), Lora, 4G, 5G, etc. to remote health centres and clouds. Extracting meaningful analyses from the raw medical data and decision making by the utilization of these extracted analyses are very powerful tools to diagnose diseases and find abnormality patterns of the patient's health status. The storage of measured and analysed data of patient's health status occurs on cloud units to be evaluated by doctors and medical staff later. For a better understanding, the basic network architecture of smart healthcare systems is visualised in Figure 1.  

Smart healthcare infrastructures interconnect the neighbour networks in a little topographical area such as a solitary specialist to numerous patient records, records of a medical clinic or assists with interconnecting gadgets at a bigger scope such as between association of medical clinics inside a city or across urban areas, the interconnection of specialists with the equivalent or distinctive specialization across a bigger locale. Information gathered through smart healthcare networks is stored in the cloud utilizing distributed computing ideas\cite{65}. Information over the cloud is processed for a smart medical services framework utilizing cognitive computing and artificial intelligence concepts. 

 The distinctive implanted body sensors, actuators proficiently gather information from the smart intuitive gadgets and sends it through 5G cell organizations or WiFi to the cloud and to the primary regulator where the tremendous measure of information is experienced careful observing the interaction. The progressed investigation in AI machines can apply to this information for infection analysis and abnormality detection. The distant observing of a patient's health status helps in lessening the length of clinic stay and forestalls reaffirmations. It additionally has a significant effect on diminishing medical care costs essentially and improving treatment results. The patient commitment as associations with specialists have gotten simpler and that is just the beginning proficient that lead to support fulfilment. On the off chance that any confusion or crisis emerges while checking of patients by information examination, consequently, an alert will trigger the smart centres like an emergency vehicle with the patient subtleties like healthcare reports, patient's careful area, conceivable essential medicines. Therefore, while the emergency vehicle will pick the patient and arrive at the clinic, the separate closest healthcare unit also will get informed about the crisis case, with the goal that the patient can ready to profit convenient medical care\cite{66,85}.

\section{Emerging Technologies in Smart Healthcare }

With the acquaintance of new technologies to improve the above-recorded individual features, the general performance of smart healthcare is improving. Each below-mentioned emerging technology provides different enhancements for different layers which can be explicated in the view of their functions and main requirements which correspond to these functions. To a better understanding, the main requirements and functions of the different layers in smart healthcare infrastructure are summarized in Figure 2.

\begin{figure}[t]
\centering
\leavevmode
\includegraphics[width=3.6in]{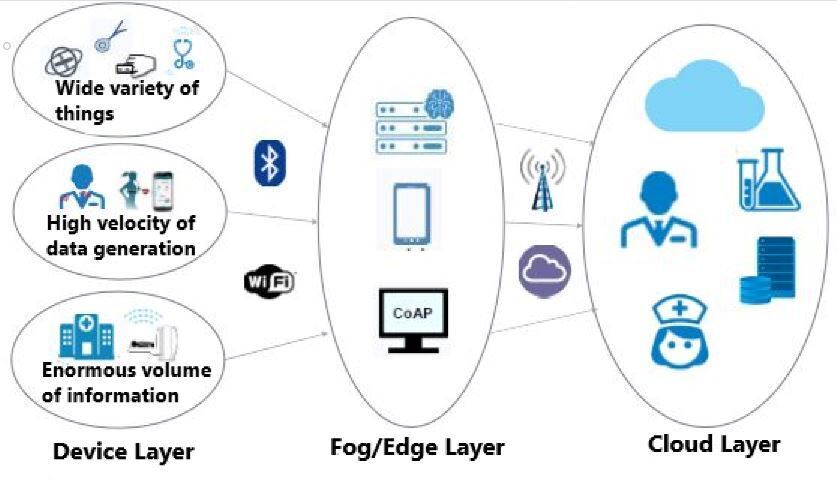}
\caption{A basic smart healthcare network architecture.\cite{11}}
\label{amplifier}
\end{figure}
\subsection{WBAN technologies}
For a better understanding of smart healthcare systems, WBANs which are the fundamental network type in IoT smart healthcare applications should be mentioned briefly. WBANs are the creating networks that are planned and produced for the human body to screen and communicate the continuous physiological boundaries. Independently associated different clinical sensors and actuators situated on, in, around or/and close to the human body constituent WBAN to screen physiological signs. A commonplace WBAN made up with one sink and a few sensor hubs on, around or embedded in the human body\cite{62}.
WBANs have an enormous potential to reform the eventual fate of medical services observing by diagnosing numerous dangerous illnesses and giving continuous patient checking and because of that, WBANs are specified for the various deployment of IoT healthcare services.
 
The main wireless communication technologies being utilized are
ZigBee, WLAN, and Bluetooth in WBANs. Also, different protocols have been proposed and intended for WBAN such as IEEE 802.15.4, IEEE 802.15.6 and IEEE 802.15.1\cite{79}. These protocols and communication technologies are selected for providing low power gadgets, low range, and low information rate.

The information gathered by body sensors such as the patient's vital sign or physiological signals such as temperature, heartbeat, and brain signals blood pressure, etc. After that, these sensors transmit the data through the wireless channel to the base
stations. The transmitted data are received by remote healthcare centres or the cloud of smart healthcare services. It is essential to have reliable, sustainable communication between the elements of smart healthcare services. In \cite{46}, The proposals and the latest solutions for refinement on the communication of smart healthcare devices are reviewed and discussed.
\subsection{Machine Learning}
ML and AI have a tremendous impact on smart healthcare systems by enhancing the management of a high volume of data, ensuring low-latency and reliable outcomes. The critical effect of AI has been on the detection and prediction of issues that necessary complex clinical tests and the utilization of ML can help in the finding of the problems continuously and give customized medical care. ML algorithms can be applied to learn the mobility patterns of the network to provide proactive solutions to the changing network dynamic. With the acceptance of AI in medical care units, the immense informational indexes created from the healthcare units can be prepared through various AI calculations to do expectation and investigation for medical services\cite{66}. It is hard for a professional or clinical expert to investigate any patient's example and indications to analyze illness from enormous data of every patient. Which can be taken care of viably by different machine and profound learning calculations with the least mistake rate and higher precision when contrasted with the medical experts. AI algorithms can deal with the immense number of data that are gathered from various smart IoT gadgets inside a group of time and anticipate the outcome, which created electronic health reports at that point ship off the separate clinical allotments for additional examination and ideas. Herewith, The utilization of ML in smart healthcare systems are in three significant zones to give customized medical services such as diagnostics, assistive systems, and patient monitoring and alarm systems. In this paper\cite{67}, the authors propose a recurrent neural network (RNN) algorithm with a long-short time-domain (LSTM) to provide energy-efficient, high accessible, predictive medical treatments. Checking the patient and quickly acting after a basic circumstance comprises significant tasks of clinical staff, in any case, a patient might be in a critical circumstance prompting his demise. First, scoring mechanism for vital signs which are held by biomedical sensors is conducted. Easing for handling the huge amount of data and providing a correlation between various medical information are achieved by this mechanism. Then the LSTM prediction method applied to the medical information which means the system can keep the necessary information for their users and predict the abnormalities by processing the historical medical data. 

DL algorithms need large datasets to provide accurate outcomes for early diagnosing high-sensitive patients who for example has signs of dementia. The necessity of large datasets leads to the consumption of more power and it should be enhanced for the real-world usage of early diagnosis mechanisms. In \cite{76}, the authors stated the current diagnosis procedures of dementia and the implementation of ML methodologies such as support vector machine, random forest, principal component analysis, naive Bayes, neural network and deep learning, on those procedures. So far studies and applications for early accurate diagnosis o dementia are discussed and the most common ML algorithms are indicated in this paper. It is also emphasised that discovering new signs of dementia and improving the current schemes and applications is challenging even in the utilization of ML and DL algorithms. Even though DL algorithms enhanced the current procedures by finding neurological signs more easily and operating speech recognition mechanisms, the current applications and studies are not sighted on the horizon for real-world utilization at least not globally-wise.

The diversity of necessary information for an efficient healthcare treatment aggravates getting the accurate health status of the patients. In \cite{81}, various ML algorithms such as Support-Vector Machines (SVM,) K-Nearest Neighbors (KNN), Decision Tree (DT) and LSTM are evaluated for human activity recognition by smartphones. Collected data of 12 different type of human activities such as walking, standing, running, etc. are compressed and processed by smartphones. ML algorithms allow classifying and clustering these unlabelled data and specifying classifiers for elderly patient ad young patients differently. According to simulation results, LSTM-based and SVM-based classifiers obtains the highest accuracy but consumes more energy than the other ML algorithms and DT-based and KNN-based classifiers provide faster results than the other ML-based classifiers.

\subsection{Fog Computing}
Fog computing means processing the measured data by sensors on the sensor devices or the devices that are closed to sensors instead of on cloud servers or remote healthcare centres. So basically, it can be said that an enhanced fog computing algorithm means a low-latency diagnosis or alert mechanism. Fog computing empowers the organization to convey the cloud administrations at the organization level with the computational capacities appropriated locally at the organization level, unlike the cloud frameworks, which are carefully concentrated. The fog nodes also help in decreasing energy use by restricting significant distance transmissions utilizing a time limit calculation for choosing the neighbours.

Fog computing and cloud computing is on-demand to be enhanced vertically with the increased amount of patients’ data. Through the low latency, high data-rated communication channels, the health data can easily transmit from one IoT node to another but processing the health data and making assumptions and decisions from the data can be still a challenging task. Computing algorithms that are mostly deep learning-based or machine learning-based needs more enhanced Graphics Processing Unit (GPU), bigger storage capacities on cloud servers or fog nodes to provide more efficient and accurate results. In \cite{71}, a smart healthcare system comprising of a pathology location framework, which is created utilizing deep learning algorithms is proposed. The pathology can be recognized from the electroencephalogram signals of a patient. In this system, a smart EEG headset catches EEG flags and sends them to a mobile fog computing server. The fog node processes the data and transmits them to a cloud computing server. The cloud computing server does the principle handling utilizing deep learning and settles on whether the subject has pathology or not. Customers and patients of the structure are associated through a confirmation administrator situated in the cloud server. Analysis results affirm the suitability of the proposed system.

In addition to the large volume of data generated by sensors and IoT devices, the measured data varies on the priority of utilization and the necessary bandwidth of transmission and this can be considered a major challenge to be solved. In \cite{78}, a bandwidth allocation procedure based on a cooperative game theory and an offloading mechanism based on non-cooperative game theory are proposed. With bandwidth allocation procedure, sensors and local IoT devices can transmit their data to edge servers based on their priority of emergency. In the proposed framework, all measured data has a timestamp and data freshness can be controlled timely. According to the necessary freshness and the priority of the data, a bandwidth threshold for each data type is decided cooperatively. The bandwidth allocation procedure can be executed to reach a minimum cost for the overall network. In the offloading mechanism, sensors and IoT devices decide periodically which edge server are used for processing and storing data to decrease network overhead and energy consumption. The simulation results indicate that the proposed framework decreases overall energy consumption and provides more low latency than the traditional algorithms. By looking at this above-mentioned work and the future need predictions of smart healthcare systems, dynamic frameworks and algorithms should be developed to cope with the large volume of different kind of medical data.

\begin{figure}
\centering
  \includegraphics[width=3.6in]{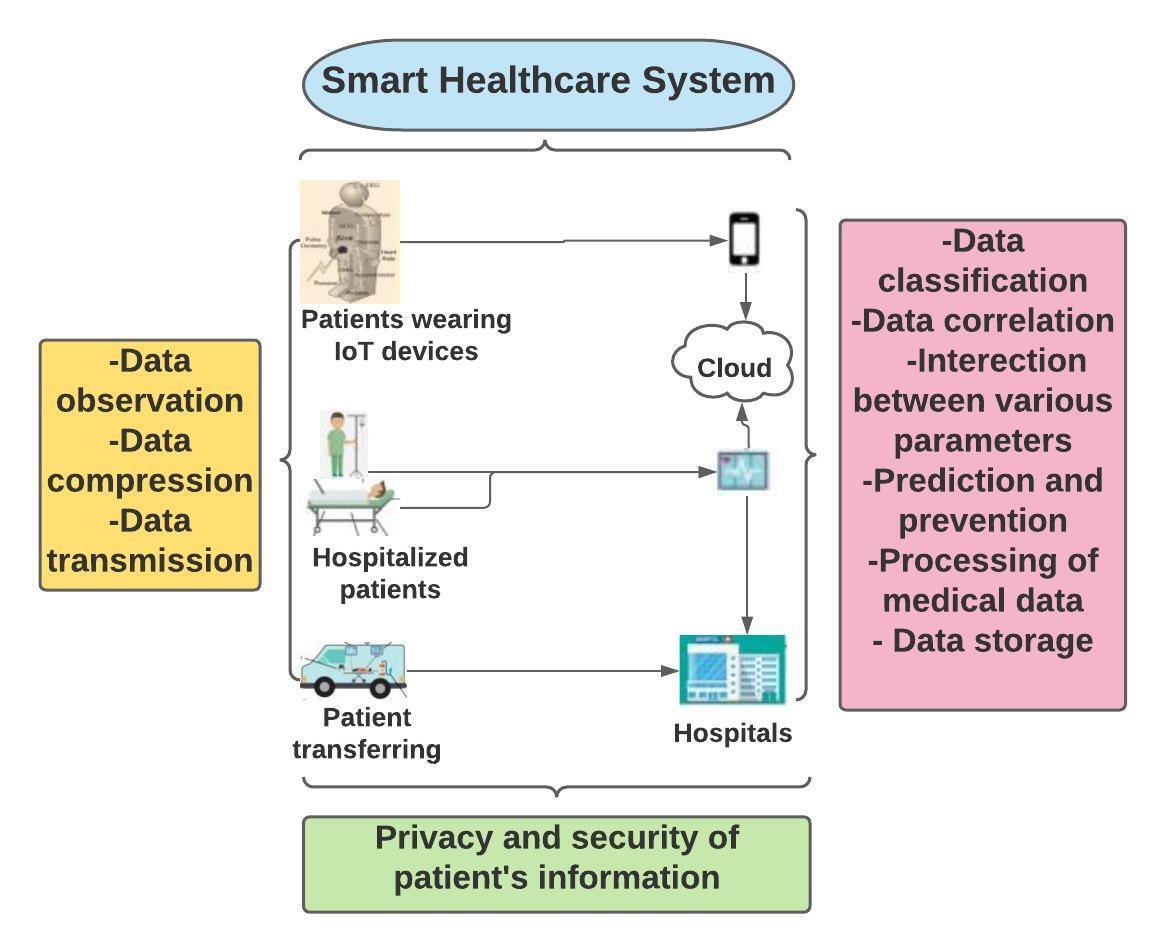}
  \caption{{The main functions/requirements of each layer in Smart Healthcare Systems}}
\label{PA_meas}
\end{figure}
\subsection{SDN-NFV technologies}
5G technologies are relied upon to empower the worldwide financial yield of 12.3 trillion by 2035\cite{48}. With technologies such as SDN and NFV, network slicing can be used to provide instant medical treatment. It is achieved by virtualization of network device functions to suit the requirements of the network in real-time such as efficient energy utilization, improved resource allocation and management, and enhanced security and privacy. Network slicing can separate the main physical layer into an unattached logical network for controlling the subject to network slice to accomplish the functions for the particular wearable gadgets[14].
SDN architecture is a key solution to make the traditional network structures less complex. SDN provides a global view of the network and a central control mechanism by separating the SDN control plane and data plane. Therefore, applications and services can be more programmable, manageable, flexible and more precise with the tools and mechanisms provided by SDN\cite{58}. Thereafter; by using NFV, network resources can be efficiently allocated to virtual networks, and adding, removing, or updating a function for all or subset of end-users becomes much more manageable. NFV guarantees the improvement of asset provisioning to the end-users with high QoS and ensures the exhibition of virtual organization activities including least latency and failure rate\cite{48}.
In \cite{77}, an SDN based smart healthcare framework with a lightweight authentication scheme is proposed. The proposed framework uses SDN technology to cope with load balancing and insufficient utilization of network resources. First, each IoT device transmits its measured sensor data to edge servers by a secured channel which is protected by hash functions and secret keys. After that, edge servers communicates an SDN controller to obtain an intelligent decision for load balancing and network resource utilization. By the centralized and intelligent control mechanism of SDN, the proposed framework can manage an efficient load distribution mechanism on the edge servers to ensure a low delayed and high-rated communication. Simulation results show that the proposed framework beats the present smart healthcare systems on aspects of latency, network overhead, throughput and average response time.

With the heterogeneous data that consists of physiological data, environmental data and health status data, the smart healthcare networks are overhead. In \cite{79}, an SDN architecture in WBANs is detailed. Each IoT devices acts as an SDN device and controlled by an SDN controller which can provide an overall view of networks and manage the network traffic. The simulation results are run on gathered data and generated data respectively. The results show that The proposed SDN architecture can provide seamless communication and low network overhead.

\subsection{Blockchain Technology}
Blockchain technology is one of the primary inventions that can be used in smart healthcare systems. Due to the content of the patients' medical data, the privacy and the integrity of the data is a major concern that must be reassured for the patients. By the decentralized and encrypted data blocks that the blockchain can provide, it can offer upgrades in decentralized capacity, circulated record, interoperability, confirmation, dependable, changelessness just as the chance to encourage secure and astoundingly successful associations between nodes such as patients, medical care suppliers, providers on the smart healthcare systems. Besides, such nodes could be dynamically and proficiently increased and managed which means blockchain technology can handle the constantly increasing number of IoT devices that can be used. In \cite{70}, a blockchain framework for smart healthcare is proposed to improve encryption and smart healthcare access control techniques and enhance the security and interoperability components to help privacy preservation. The proposed system utilizes a private blockchain swell chain to build up reliable correspondence by approving nodes dependent on their operable construction. So that, controlled correspondence needed to settle combination and coordination issues are encouraged through various nodes of smart healthcare infrastructure. Due to the nature of Blockchain technology, each newly generated block is added to the blockchain and that can be impractical considering the amount of medical data in smart healthcare systems. In \cite{74}, A system based on blockchain named Healthchain is proposed. In this model, symmetric keys are used for decryption and encryption of the medical data and diagnosis on the user node and doctor node respectively. The medical data which is generated by the user node is stored on the various servers to reduce the amount of the data in the blockchain and only the hash strings of the medical data is added to the blockchain. Besides, traditional security schemes update the privacy of the user each time a user makes adjustments to the medical data. Instead of that, the proposed model offers users an opportunity of updating their keys independently with the medical data. Simulation results indicate that the proposed model beats the traditional security schemes at providing low communication and storage costs and a more reliable and secured system for their users.

\section{Applications and Services of Smart Healthcare}

In recent years with emerging IoT technologies in the healthcare area, many applications and services start to be used extensively in our daily lives. Each application or service is designed to provide increased patient satisfaction and better self-management. Some of the main applications and services in smart healthcare systems are presented in this section, and also in Table II.

\subsection {Applications}

\par

\subsubsection{Electrocardiogram Monitoring (ECG)}

An electrocardiogram (ECG) is a straightforward test that records the heart's electrical action. A standard heart's movement checking by a handheld ECG gadget is both compelling and gives long-term cost investment funds to heart patients. Non-contact, capacitive-coupled anodes are likewise usually utilized for ECG signal securing as they don't require direct contact with the skin, empowering their utilization as wearable sensors implanted in dress \cite{9}.
In \cite{82}, a real-time ECG monitoring system is proposed. The collected ECG signals can be processed and transmits to edge servers and cloud servers. With an android-based application, doctors and patients can access to ECG information of the patients remotely. Simulation results shows that this proposed application can provide a reliable and low-latency interface for the ECG of the patients.

\subsubsection{Blood Pressure Monitoring}
Hypertension is a genuine hazard factor that may prompt coronary episodes or strokes. Blood flow may differ from minute to minute and continuous monitoring of blood pressure by medical sensors helps the patients for being able to know their health's status. In \cite{87}, the blood pressure measurement and calibration system are proposed. The proposed system has two ends namely as sensing-end and server-end. The sensing end consists of wearable sensors which can monitor blood pressure and Internal Measurement Unit (IMU) that can provide a correlation between sensed data and its physical movement. The server-end provides management of network function, processing and analyzing the raw sensed data. The sensed blood pressure data and IMU data can be calibrated by extracting features and constructing matrix vectors of these features. The system utilizes a regressing algorithm such as KNN, SVM, Bayesian, DT respectively to predict the future values of blood pressure measurements. The regression process submits a root means square error of the raw data. At the same time, users and doctors can access the visualization of data analysis and working status of the system on the web application of the server system. Simulation results show that the root mean square values of the original blood pressure data is increased by the SVM regression model and Decision Tree regression model. Therefore, The proposed system has good results in analyzing the sensed data and predicting the future values of the original data. 

\subsubsection{Body Temperature Monitoring}

In the detection of many viral diseases, high body temperature is considered to be one of the early symptoms of these diseases. Doctors can provide more effective treatment in case of getting instant and accurate alerts by real-time monitoring and body sensors when the body temperature of patients get out of a certain range.

\subsubsection{Rehabilitation system}

Physical medication and restoration can upgrade, what's more, reestablish the practical capacity and personal satisfaction of those with some physical weakness or handicap. The IoT helps all viable remote meetings in exhaustive rehabilitation. Therefore, a smart rehabilitation system fills a crucial part of medication. For example, ın this paper\cite{86}, a smart framework for home rehabilitation of post-stroke patients is proposed. In this framework called SmartPants, four biosensors are attached to the patient's body with plastic bands and a pressure sensor attached to the patient's feet to understand the distribution of patients weights. With these sensors, the framework can monitor that a patient does the exercises accurately and gives an alert sign to the patients for inaccurate exercising. The proposed system can measure and process the sensed raw data in the context of five of the most common exercise types for post-stroke patients. After that, the framework can classify the 64 features that are obtained from the sensors by the ML algorithms such as Random Forest, Random Tree, Naive Bayes and Multilayer Perceptron. The simulation results show that the recognition and classification of complex movements cause more inaccuracy than simple movements. On the other hand, the framework can provide good overall accuracy of around 96.5 percent independent of the ML classifier.

\begin{table}[ht]
    \centering
\begin{tabular}{ |p{3.3cm}||p{3.5cm}|}
 \hline
 \multicolumn{2}{|c|}{IoT Healthcare Systems} \\
 \hline
Applications & Services\\ 

 \hline
 
• Glucose Level Sensing  &  • Adverse Drug Reaction (ADR)     \\
• Body Temperature Monitoring  &  • Ambient Assisted Living(AAL)\\
• Blood Pressure Monitoring  &  • Mobile Internet of Things(m-IoT)\\
• Electrocardiogram Monitoring (ECG)  &  • Community Healthcare(CH)\\
• Oxygen Saturation Monitoring &  • Children Health Information (CHI)\\
• Wheelchair Management  & \\
• Rehabilition System  & \\
• Medication Management  & \\

 \hline
\end{tabular}
\caption{The applications and services of smart healthcare systems}
\label{table:ta}
\end{table}

\subsection {Services}

\subsubsection{Ambient assisted living}

Ambient assisted living (AAL) is a smart healthcare service for elderly people who need continuous care in their homes. The main purpose of AAL is to expand the self-sufficient presence of more seasoned individuals in the homes favourably and securely. AAL can provide an independent lifestyle for elderly patients. Although the current home-based AAL care systems for elderly people offer very sufficient results, most of these systems have little contribution to providing mobility for their users. One of main causes of the above-mentioned issue can be considered as the turned off mobile phones because elderly people may forget to charge their mobile phones. Another cause is that if communication is lost between an IoT device and its assigned gateway, a short-range protocol cannot establish a new communication and transmit the medical data to an unknown gateway. In \cite{88} , an emergency situation detection system for AAL is proposed. A BLE-based IoT sensor can monitor the heart-rate data and fall detection to transmit to a third party mobile relay. a mobile device which is called a donor in this system must be located in proximity to transmit the sensed data to a cloud server. Mobile relays, which are selected based on their Received Signal Strength Indicator (RSSI) values, are not responsible for processing the sensed data but only transmitted to a cloud server. Each IoT sensor can store the measured data in their storage module for the amount of this space. After the processing part, a server can send an SMS or e-mail to carers who are responsible for the patients if an abnormality has detected from the sensed data. 

\subsubsection{Internet of m-health Things (m-IoT)}

Physiological signs of the human body are gathered by methods for various wearable sensors by the product application. Depending on the situation, signals in the structure of short messages can inform medical care experts about health-related crisis foundations and help them in taking the appropriate activities. Immediate access to the medical data offers more chances to enhance the quality of human services, improves persistent fulfilment, and supports convenient mediation \cite{ahmadi2019application}.

\subsubsection{Adverse drug reaction}
Adverse drug reaction is a condition that is related to taking medicines. With the assistance of a pharmaceutical insightful database, the medical data is then planned to detect whether the medication is perfect with its hypersensitivity profile and the health database \cite{25}. counterfeit drugs in the pharmaceutical supply chain is a big challenge to provide sufficient medicine-based healthcare. In \cite{89}, Blockchain-based IoT implementations on medicine management are discussed for a secure and controllable drug supply chain. The integrity of the drugs can be corrupted in four stages. Due to forgery at the stages of the ingredient's attainment and the production of the drug,  the produced drugs may not provide efficacy as needed. At the stage of the distribution, drug delivery is difficult to trace especially in borders of third-world countries. And at the pharmacy stage, the black market and online sales of drugs cause the utilization of insufficient medicines by users. The transparency of Blockchain technology and the traceability of IoT systems offer a secure and visible pharmaceutical supply chain to all participants of today's pharmaceutical industry.

\subsubsection{Community healthcare}
The widespread of IoT devices will bring opportunities in healthcare systems such as remote monitoring of patients and providing instant alert mechanisms. The benefits of smart healthcare systems provide enhancements for not just individuals in urban places but also individuals in rural places where may be difficult to access proper and efficient healthcare. In \cite{101}, an e-health monitoring system is proposed to reduce hospital's visits and time-consuming processes. The proposed model collects the individual's medical data related to temperature, heartbeat rate, blood pressure and transmits it to a healthcare expert for the betterment of the rural communities. Community healthcare means the utilization of medical data and analysis from a broadband perspective. Especially in crisis such as floods, earthquakes, cyclones, and epidemic breakouts, there is a huge need for an overall analysis of individuals to take specific and efficient actions. For example, in \cite{100}, a rapid response plan for smart cities is introduced. The monitoring unit which consists of a sensor and a fog node can monitor the vital signs of the patients and process the sensed data in the proposed model. And also, this unit can transmit the data to node clustering units which can group the nodes in similar locations and compare them with optimal parameters. The grouped data are transmitted from the source to the destination through the shortest path which calculates in the path designer unit. Path designer unit is implemented in cloud servers and provides communication between a patient and the nearest healthcare facility. And also, the alert mechanism which is implemented in the path designer unit can give alerts such as critical, warning and safe to a user interface. Simulation results indicate that this proposed model is very efficient to reduce network costs.

\section{The Challenges of Smart Healthcare}
Designers should be aware of the demands and real-life usage of smart healthcare systems to fulfil the requirements and overcome the challenges of smart healthcare. The challenge descriptions and the proposals of the researched papers which are discussed in this section are summarized in Table III. And also, the main challenges and benefits of smart healthcare systems are mentioned briefly in Table IV. 
\subsection{Large Volume of Data }
One of the main challenges of smart healthcare is the management of a very huge amount of medical data. These data are produced by IoT sensors that attached to human bodies and because of that, there is continuously changing data of the patient's health status. Also, with the rising need for detailed health analyses, different kinds of medical data with different kinds of formats are created on sensors and access points. 
In \cite{34}, a wearable monitoring node gathers the ECG information, and using WiFi technology is transmitted directly to the IoT cloud and is stored on the database for offline storage. At the point when any distortion is discovered, a programmed email is shipped off the clients furthermore, specialists for dissecting the basic states of the patients and gives crisis health assistance. In \cite{35}, an online medical decision support system is introduced for chronic kidney disease prediction and the entire experimental process clearly pointed out that the presented model offers excellent classification with the maximum predictive accuracy of 97.75 on the tested dataset. In \cite{36}, the authors propose an adaptable three-level framework to store and process such a huge volume of wearable sensor information. Level 1 focuses on the assortment of information from IoT wearable sensor gadgets, level 2 deals with storing the huge volume of wearable IoT sensor information in cloud computing, and level 3  deals with building up the strategic regression-based expectation model for heart sicknesses.

The medical data gets bigger, the variety of the medical data is also affected and increases. Mostly, a healthcare diagnosis and treatment require various kind of medical data. Traditional processing algorithms can have difficulties ensuring a proper result for the patients and at this point, DL algorithms can be life-saver settlements. In \cite{72}, a deep neural network (DNN) model is proposed for predicting depression risk using multiple regression. The setting of the proposed DNN model consists of the data to foresee circumstances and conditions affecting depression in light of setting data. Every setting data identified with indicator factors of depression turns into a contribution of DNN, furthermore, factor for depression expectation turns into a result of DNN. For DNN association, the regression examination to foresee the danger of sadness is utilized to anticipate the potential setting affecting the danger of depression. As indicated by the presentation assessment, the proposed model was assessed to have the best execution in regression investigation and similar examination with DNN.

\subsection{High-Power Consumption}
The demands from users are in the directions of executing multiple tasks such as data gathering, pre-processing, and real-time transmitting data and because of these reasons, devices must use high power. Correspondingly, the consistent need for charging devices can be a burden for most users especially elderly patients. To solve this challenge, researchers address to put devices on sleep mode or power-efficient mode at unnecessary periods of monitoring or high sophisticated algorithms for reducing power consumption of smart healthcare devices. In \cite{37}, authors present a novel real-time quasi-random signal (QRS) detector and ECG compression architecture for energy-constrained IoT healthcare wearable devices. A lossless compression procedure is incorporated into the proposed 
design that utilization the ECG signal first subordinate, variable-length encoder. Pressure design helps IoT clinical gadgets to accomplish super low power activity and limit the information that should have been sent to limit power utilization for gadgets outfitted with remote transmitters. In \cite{42}, it is proposed that a light-weight ECG signal quality assessment (ECG-SQA) method for automatically assessing the quality of acquired ECG signals under resting, ambulatory, and physical activity environments. Thereby, this method is used for improving the battery life of IoT-empowered wearable gadgets and decreasing the cloud server traffic burden, data transfer capacity, and treatment costs. In \cite{46}, a novel self-adaptive power control-based enhanced efficient-aware approach (EEA) is proposed to reduce energy consumption and enhance battery lifetime and reliability. The proposed EEA and conventional constant transmission power control (TPC) are evaluated by adopting real-time data traces of the activities and cardiac images. This method uses various network parameters such as targeted RSSI, path loss, the wireless channel coefficients to provide power level by the battery need of receiver nodes. In \cite{83}, a cluster-based model for smart healthcare systems is proposed. The proposed model utilizes the neighbour sensors and IoT devices as a cluster and appoints a cluster head to each cluster. Cluster heads can communicate with each other and selects a leader to transmit sensed data to a base station. Simulation results indicate that the proposed model provides more lifetime to the smart healthcare networks than the traditional models.

\subsection{Security and Privacy}
With the accommodation of smart healthcare devices into our daily lives, the IoT infrastructure of healthcare must ensure the security of patient's private information, the integrity of medical data, data freshness and authorized access to information, etc. to complete user's satisfaction. According to researches, many security attacks may occur at any layer of communication protocol. In \cite{17}, the security threats such as confidentiality, authentication, privacy, access control, trust, and policy enforcement of IoT systems are analyzed and the existing security algorithms and techniques such as Advanced Encryption Standard, Data Encryption Standard, Rivest-Shamir-Adleman are compared for addressing the security challenges. If medication information of patients is accessed by hackers and falls into the wrong hands, the treatment of patients can be influenced badly. In \cite{41}, it is introduced that the noise-aware biometric quantization framework (NA-IOMBA) capable of generating unique, reliable, and high entropy keys with low enrollment times and costs with several experiments. In \cite{43}, it is proposed that a reinforcement learning-based offloading algorithm for an IoT device achieves the optimal offloading policy via trial-and-error without being aware of the privacy leakage, IoT energy consumption, and edge computation model. In \cite{45}, the authors propose a secure system to devise a novel two-fold access control mechanism, which is self-adaptive for both normal and emergency situations. Besides all of these, the blockchain is contributing to securing these systems by providing a transparent system of data storage and leveraging smart contracts to secure the services. In the blockchain technique, every patient has the power to conceal their own data from different substances or individuals. The patient has the power to allow different people to see their reports whatever the patient needs to show them. Inside the blockchain innovation, at whatever point exchange or a test, for example, medication is prescribed to a patient, different elements may approve the test, medication cost or generally cost of treatment relating to specific infections has been reflected into the chain. The specialist or some other middle of the road or outsider may not profit the patient relating to pointless tests proposal to cover rock-solid bills of medicines, undesirable medication costs. In \cite{50}, it is given that a security structure of medical services media information through blockchain procedure by creating the hash value that is provided to the entities databases so that any change or adjustment in information or breaking of drugs might be reflected in whole blockchain network clients. The proposed system uses blockchain technology for guaranteeing the security and transparency of the patient's record, report accessibility, and shipment process among supplier and client.

\subsection{Low-latency Tolerance}
  Working with such heterogeneous and significant amounts of data may obstruct reaching these data for any time and unless continuous access to information is provided, consequences may be very deadly for some critical patients and critical situations. Critical situations such as reduced blood pressure or rapid changing of heart rate or just simply fall of an elderly person remote monitoring healthcare devices must alarm the patients' doctor or their caretaker. In \cite{38}, this paper presents a truthful and efficient mechanism that can prevent gateways from strategically misreporting the priority levels of medical packets, while incentivizing the base station to manage the transmission scheduling according to the desired manner. In \cite{44}, a reduced variable neighbourhood search-based sensor data processing framework is proposed to enhance the reliability of data transmission and processing speed. This reliable transmission mechanism can recollect lost or inaccurate data automatically. According to the simulation results, this mechanism can improve the successfully delivered ratio as well as optimize the resource allocation.
\begin{table}[ht]
    \centering
\begin{tabular}{ |p{3.4cm}||p{3.3cm}|}
 \hline
 \multicolumn{2}{|c|}{IoT Healthcare Systems} \\
 \hline
Challenges & Benefits\\ 
 \hline
• Scalability and Interoperability &  • Wide availability     \\
• The high power consumption  &  • Instant and reliable treatment\\
• Low latency tolerance &  • Cost reduction\\
• Necessity of user-friendly devices  & • Effective medicine and disease control\\
• Security and Privacy &  • Easy Usage\\
• Computational Intensity  & • Improved communication\\
 \hline
\end{tabular}
\caption{The main challenges and benefits of smart healthcare systems}
\label{table:ta}
\end{table}

 Providing an understandable, simple inference of healthcare status can be difficult because of the amount and the variation of the medical data. In \cite{39}, authors present an IoT based smart edge framework for remote monitoring, in which wearable sensors send information into a novel programming motor that changes voluminous sensor information into clinically important results. the proposed architecture can provide a low-latency transmission and readable data visualization. The implementation occurs in a super-speciality hospital and a rural region that is 300 km far away from the hospital. The collected data by body sensors are transmitted to the edge servers or the cloud. After the normalization, the results are specified by the patient’s medical records and the target disease. The outcomes of this proposed work are compared to the diagnosis of the real doctors and health workers and the proposed smart healthcare system has very efficient outcomes on providing readable visualization of medical information. In \cite{57}, authors submit an android based framework that consists of data collection and data visualization modules, and the application development used in this system is to build a user-friendly interface to know the status of a patient by a medical expert. 

\subsection{Interoperability and Scalability}
 Standardization of communication technologies for wearable gadgets and implantable sensors to give consistent availability is critical\cite{61}. Due to the various features of smart healthcare devices, many different communication protocols are used to fulfil the requirements of these devices. For example, in the USA the standardization of wireless medical devices demands a multiagency regulatory environment, involving three agencies; Food and Drug Administration (FDA), Centers for Medicare and Medicaid Services (CMMs), and Federal Communications Commission (FCC) \cite{11}. In \cite{56}, authors have proposed the semantic interoperability model for big-data in IoT (SIMB-IoT) to convey semantic interoperability among heterogeneous IoT gadgets in the medical services area. This model is used to recommend medicine with side effects for different symptoms collected from heterogeneous IoT sensors.
 
 Each IoT healthcare service manufacturer uses different platforms to provide communication between their products. In the context of solving this interoperability issue, the INTER-IoT European project is introduced in 2020 and INTER-Health framework has been build upon on this project to provide interoperable, user-friendly healthcare services.
 In \cite{63}; the proposed INTER-Health platform means to build up a coordinated IoT framework for observing people's way of life in a decentralized portable approach to forestall medical problems coming about from food and active work problems. This observing cycle can be decentralized from the medical services community to the observed subjects' homes and upheld in portability by utilizing on-body physical activity screens. The proposed platform can be deployed over two specific platforms named universAAL and BodyCloud. In particular, the computerized observing at the medical services place would be upheld by universAAL, while the physical activity checking in versatility would be empowered by the BodyCloud versatile administrations. At last, a particular programming application, comprising of a professional web tool (PWT), is additionally coordinated inside the correspondence design all together to give all the wellbeing administrations in an easy to understand route to the specialists. The results from different use cases demonstrate this proposed platform provides user-friendly monitoring, accessibility to a high number of users and a high continuation rate of users.  

\begin{table*}[t!]
\caption{Literature paper research.}
\centering

\begin{tabularx}{\textwidth}{ p{1.2cm} p{4cm} p{2.cm} p{6cm} p{1.2cm} }.\\

\ Challenges &Description  & Consequences
/Effects&Proposal& Ref \\

  \toprule
  
   Computational intensity &
 
 • Loss of information.
 
 • Restricted storage and computation resources exist in mobile gadgets.
 
 • Inability of the traditional data processing methods to store a huge volume of data.
 
 • The difficulty of providing an accurate correlation between various medical information.
 
 • The network overhead because of the heterogeneous data.
 &
 • Continuously changing data of the patient's health status
 & • When any abnormality is discovered, a programmed email is shipped off the clients, otherwise, the medical data can be stored in an SD Card. 

 • An online decision support system should be used for disease prediction and alerts.
 
 • Novel architectures should be considered to store and process such a huge volume of wearable sensor data.
 
  • A LSTM-based algorithm predicts abnormalities by processing various medical information.
  
  • An SDN architecture in WBANs can enable an overall view of networks and manage the network traffic to provide seamless communication and low network overhead.
  &
 \hspace{1sp}\cite{34}, \cite{35}, \cite{36}, \cite{67}, \cite{79}
 \\
 \midrule

High power consumption &
  • In order to provide real-time monitoring efficiently, devices and nodes must be charged with sufficient energy. 
  
  • High power consumption of smart devices and shutting down of the devices may cause vital consequences.
  
   • In order to execute multiple tasks by smart devices, they must use high power.&
   
  • Limited power batteries for prolonged time periods.

  &

  • The performance of the sensor node can be considered before deciding on the amount of energy to be given.
  
  • End-devices should control their energy usage to enhance the battery lifetime by power control algorithms.

  • Compression architectures can be considered for energy-constrained IoT healthcare wearable devices.
  &
\hspace{1sp}\cite{31}, \cite{46},  \cite{37},
  \\
  \midrule
  
  Security and privacy &
  • Incorrect patient identification.
  
  • Easily cracked processes and unreliable medical information.
  
  • An attacker can estimate the size of the sensing data newly generated and thus evaluate the usage pattern.
  
  • The onsite first-aid personnel are not permitted to get the patient’s historical medical data.
  &
  • Violation of patient's private information, unreliable medical data, and unauthorized access to information.
 &
 • A noise-aware biometric framework is capable of generating unique, reliable, and high entropy keys with low enrollment times and costs with several experiments. 
  
 • A continuous security solution based on IoT using biometrics for smart healthcare technologies. 
   
 • A RL-based offloading algorithm for an IoT device to achieve the optimal offloading policy via trialand-error without being aware of the privacy leakage, IoT energy consumption, and edge computation model.
 
 • A secure system to devise a novel two-fold access control mechanism, which is self-adaptive for both normal and emergency situations.
  &
\hspace{1sp}\cite{41}, \cite{33}, \cite{43}, \hspace{0.05cm}\cite{45},
 \\
 \midrule
 
Low latency tolerance &
  • The accuracy of the patients' health status and instant treatment. 
  
  • The accessibility of medical information and instant diagnosis.
  
  • Inefficient movement, loss of balance, and inability to walk
  
  • Receiving late results.
  &
   • Late or incorrect diagnosis. &
  •  Online health status prediction models can be used for instant diagnosis.
  
  • Truthful and efficient data-processing mechanisms can prevent misreporting the priority levels of medical packets and manage the transmission scheduling according to the desired manner.
  
  • It can be used monitoring systems that can recognize a fall, and then automatically send alerts.
  
  • Data processing frameworks can be used to enhance the reliability of data transmission and processing speed.
  &
 \hspace{1sp}\cite{32}, \cite{38}, \cite{40}, \hspace{0.05cm}\cite{44},
 \\
 \midrule

The necessity of user-friendly smart devices &
  • Difficulty of providing an understandable, simple inference of healthcare status because of the amount and the variation of the medical data. 
  
  • Necessity of assisting medical experts to view the graph of any particular parameter and analyze the data to predict and conclude a concrete decision&
  
  • The lack of availability and accessibility of the medical data by medical workers. &
  
  • Software engines should use to transform enormous sensor data into clinically meaningful summaries.
  
  • Applications that consist of data collection, data visualization modules, and build user-friendly interface to know the status of a patient by a medical expert&
 \hspace{1sp}\cite{39}, \cite{57},
 \\
 \midrule
 
 Interoperability and scalability&
 
 • Extremely heterogeneous IoT devices.
 
 • The delay or variation in the data analysis.

 &
  • The very huge amount of IoT devices.
 &
 
  •An intelligent health cloud can provide semantic interoperability to data collected from IoT devices.
  
  • Data interoperability methods can provide the meaning of data from heterogeneous sources while bringing uniformity for the data format.&
   
  \hspace{1sp}\cite{56}, \cite{60},\\

 \bottomrule 
 &
   &
   &
   &
   \\
 &
   &
   &
   &
  
\end{tabularx}  
\end{table*}

\section {Future Trends in Smart Healthcare}

Along with the challenges that are mentioned in the previous section, the relevant studies and the interests of governments and giant-tech companies such as Intel, Google, IBM, Microsoft, Apple, etc. are rising to solve these challenges. The stability among diminishing the expense of care and conceivably improving influenced individual results weighed against influenced individual protection, information privacy, and digital security dangers might be an imperative thing in the broad spread of smart healthcare.\cite{13}.

Wearable devices and smart clothes can be called the backbones of smart healthcare for non-ill persons who want to check their health status as long as they provide unplugged services for a reasonable period. Communication standards will persistently develop to supply higher data rates with diminishing power consumption while IoT technologies will profit from future very large scale integration(VLSI) technologies that need lower battery power in their activities \cite{14}. 

And also, one of the important features of smart healthcare devices is being easy to carry and new technologies such as micro-electro-mechanical systems (MEMS) can help this feature for providing constant data generation. With the improvement of MEMS-based sensors, shrewd materials, brilliant textures, and novel bio-materials alongside far off checking, what's more, the disease conclusion is a floating exploration zone\cite{61}. The advances in the MEMS is prompting a dramatic increment in the scope of sensor hubs accessible for checking the indispensable boundaries of the human body.

With the opportunities of the emerging 5G technologies, smart healthcare systems will be more satisfied by the end-users. At the point of providing well-satisfied smart healthcare, technological trends of 5G such as massive Multiple Input Multiple Output, SDN, NFV, extremely small cells, mm-wave communication, M2M communication have importance to come through the above-mentioned challenges. Especially the usage of SDNs is moving IoT from application-explicit frameworks to a more programmable biological system. The assistance of SDN makes the reasonability of the entire organization simple and wipes out the manual order line interface. The future smart healthcare networks are relied upon to be a blend of 5G and IoT gadgets which are relied upon to increment cell inclusion, network execution, and address security-related concerns. For example, in \cite{3}, it was introduced that the detailed audit of the 5G innovation and its rising job in IoMT for changing the whole scene of the healthcare industry by diminishing the size of the sensor hubs. The upsides of blockchain in smart healthcare are not restricted to make sure about sharing of patient information over stages, upgrading interoperability of information, and expulsion of outsiders for access control.

To overcome the aforementioned challenges and provide much more effective services than the traditional utilization of IoT devices, the ML field is waiting for integration into IoT healthcare frameworks. The unmatched biomedical and motion data ensure limited outcomes for the user and the medical staff. Machine learning techniques such as deep learning and reinforcement learning can be used to label the user' heterogeneous data. In \cite{64}, the authors introduced deep learning improved activity recognition in IoT conditions. A semi-supervised deep learning structure is planned and worked for more exact motion recognition, which productively utilizes and dissect the feebly named sensor information to prepare the classifier learning model. To more readily take care of the issue of deficiently named test, an astute auto-marking plan dependent on Deep Q-Network is created with a recently planned distance-based prize principle, which can improve the learning effectiveness in IoT conditions. A multi-sensor based information combination component is then evolved to consistently incorporate the on-body sensor information, setting sensor information, and individual profile information together, and an LSTM-based arrangement technique is proposed to recognize fine-grained designs as indicated by the significant level highlights logically removed from successive motion information.

Early diagnosing critical diseases such as Alzheimer's disease, brain, lung and breast cancers that can affect the patients effectively is considered to be a crucial feature for healthcare systems. Besides traditional healthcare services, smart healthcare innovations can enhance providing this feature progressively but they need ML algorithms for accurate classification and diagnosis. In \cite{73}, the authors proposed a DL-based model for feature selection and image classification. The proposed model has three main stages such as pre-processing, feature selection and image classification respectively. In the pre-processing stage, the algorithm process the raw medical image data to utilizable data for the later stages. In the feature selection process, the algorithm selects only the features that are necessary for diagnosing specific above-mentioned diseases. In the image classification stage, the algorithm decides that if the outcomes of the previous stages indicates are malignant or benign. The results show that the algorithm provides a high-rated accuracy. ML and DL algorithms should be considered as the first reference in healthcare systems for providing accurate early diagnosis of critical diseases in the future. 

  At present, novel schemes are being introduced to finding a novel and functional answer for the early discovery of the connected neurological indications. As indicated in this paper\cite{68}, there is a solid relationship between the beginning of these neurological illnesses and outer side effects like the kind of eye blinking action and a novel scheme is introduced. A significant benefit of this scheme regarding the cutting edge lies in the way that this execution gives a straightforward approach to persistently screen commonplace indications identified with basic neurological illnesses. This methodology uses the solace and effectiveness of utilization natural in wearing conventional glasses while preparing the outline with helpful sensors hence empowering clever abilities on this gadget. Results show that the employed solutions can productively perform eye blinking discovery, accomplishing tantamount exactness regarding proficient clinical instruments, and giving to the patients at a similar time the advantage of compactness, solace and ease of utilization. In this context, more solutions should be utilized for the early diagnosis of neurological diseases in the future.

Blockchain technology should be considered a huge opportunity for providing secure and reliable smart healthcare systems. The sensitivity of the medical data is very high because of the content of the medical data which means the protection of the integrity and the accessibility of the medical data by the authorized users requires development by intelligent security systems. Blockchain-based security schemes should be developed and integrated into smart healthcare systems for a more reliable and secured healthcare experience.

The effects of the Covid-19 pandemic have brought the issues into open such as the importance of protection of the individuals and the low-latency healthcare services. Besides, there is a huge necessity of the large amount of the medical data which may be related to Covid-19 disease and DL algorithms and blockchain technology can be easily implementable solutions to the current systems. In \cite{75}, a smart healthcare framework based on DL and blockchain technology with the utilization of beyond 5g and 6g communication is proposed. The proposed framework has three stages namely user stage, edge stage and cloud stage respectively. The medical information collected by body sensors or in-home gadgets at the user stage such as cough signals, eye signals, body temperature signals, etc., can be transmitted to edge servers which is close to the hospitals or medical centres. The edge nodes process the various kind of gathered medical data of users and the data which is collected by CT scans and X-rays. After that, edge nodes use DL algorithms such as ResNet50, deep tree, and Inception v3 to make classification and predictions for a low-latency diagnosis and alerts for their users. At the cloud stage, DL algorithms and blockchain technology is utilized to make accurate decisions on the large data which is collected and stored from different regions and countries. In this framework, it is also enabled that a mass surveillance system and a smartphone app for their users to make decisions on social distancing, mask-wearing and risky areas for Covid-19 disease. This framework indicates that the utilization of beyond 5g and 6g communication systems, DL algorithms and blockchain technology can be very useful weapons in the battle to pandemics such as Covid-19 pandemic and the hardware suppliers and the software designers which work on smart healthcare systems should develop and enhance frameworks and schemes such as this proposed one.

\section {Conclusion}

 In the grand scheme of smart healthcare, its main objective is the high deployment of smart healthcare devices by users in their daily lives. Besides all of the benefits of the smart healthcare system, it has its issues such as interoperability, energy, security, resource management, low latency tolerance, etc. and along with the development of IoT technologies. Researchers are in a tendency to provide better healthcare service by introducing new methods and techniques to overcome the challenges in section IV. Therefore, many emerging technologies such as machine learning, AI, blockchain, fog computing, SDN, NFV, network slicing, etc. are in demand to develop smart healthcare systems. In this paper, we discuss the major emerging technologies in use for smart healthcare. We mention the main applications and services of smart healthcare systems. We highlight the challenges that IoT healthcare systems deal with and we sort and stated the latest methods and techniques that the authors have proposed. With all the studies cumulatively, smart healthcare is proceeding and will be proceeding to spread to our daily lives.

% ====== REFERENCE SECTION

\bibliographystyle{unsrt}
\bibliography{references.bib}

\begin{thebibliography}{10}

\bibitem{1}
Hossein Ahmadi, Goli Arji, Leila Shahmoradi, Reza Safdari, Mehrbakhsh Nilashi,
  and Mojtaba Alizadeh.
\newblock The application of internet of things in healthcare: a systematic
  literature review and classification.
\newblock {\em Universal Access in the Information Society}, 18(4):837--869,
  2019.

\bibitem{3}
A.~Ahad, M.~Tahir, and K.~A. Yau.
\newblock 5g-based smart healthcare network: Architecture, taxonomy, challenges
  and future research directions.
\newblock {\em IEEE Access}, 7:100747--100762, 2019.

\bibitem{69}
V.~P. {Darcini S.}, D.~P. {Isravel}, and S.~{Silas}.
\newblock A comprehensive review on the emerging iot-cloud based technologies
  for smart healthcare.
\newblock In {\em 2020 6th International Conference on Advanced Computing and
  Communication Systems (ICACCS)}, pages 606--611, 2020.

\bibitem{52}
Lamia Fourati and Sana Said.
\newblock {\em Remote Health Monitoring Systems Based on Bluetooth Low Energy
  (BLE) Communication Systems}, pages 41--54.
\newblock 06 2020.

\bibitem{51}
Imen Ben~Ida, Moez Balti, Sondès Chabaane, and Abderrazek Jemai.
\newblock {\em Self-adaptative Early Warning Scoring System for Smart
  Hospital}, pages 16--27.
\newblock 06 2020.

\bibitem{54}
Shreshth Tuli, Nipam Basumatary, Sukhpal~Singh Gill, Mohsen Kahani,
  Rajesh~Chand Arya, Gurpreet~Singh Wander, and Rajkumar Buyya.
\newblock Healthfog: An ensemble deep learning based smart healthcare system
  for automatic diagnosis of heart diseases in integrated iot and fog computing
  environments.
\newblock {\em Future Generation Computer Systems}, 104:187 -- 200, 2020.

\bibitem{49}
A.~{Ramachandran}, A.~{R.}, P.~{Pahwa}, and A.~{K.R.}
\newblock Machine learning-based techniques for fall detection in geriatric
  healthcare systems.
\newblock In {\em 2018 9th International Conference on Information Technology
  in Medicine and Education (ITME)}, pages 232--237, 2018.

\bibitem{95}
Jun Qi, Po~Yang, Geyong Min, Oliver Amft, Feng Dong, and Lida Xu.
\newblock Advanced internet of things for personalised healthcare systems: A
  survey.
\newblock {\em Pervasive and Mobile Computing}, 41:132--149, 2017.

\bibitem{93}
M.~M. {Alam}, H.~{Malik}, M.~I. {Khan}, T.~{Pardy}, A.~{Kuusik}, and Y.~{Le
  Moullec}.
\newblock A survey on the roles of communication technologies in iot-based
  personalized healthcare applications.
\newblock {\em IEEE Access}, 6:36611--36631, 2018.

\bibitem{98}
Yasmeen Shaikh, V.~K. Parvati, and S.~R. Biradar.
\newblock Survey of smart healthcare systems using internet of things (iot) :
  (invited paper).
\newblock In {\em 2018 International Conference on Communication, Computing and
  Internet of Things (IC3IoT)}, pages 508--513, 2018.

\bibitem{97}
Hossein Ahmadi, Goli Arji, Leila Shahmoradi, Reza Safdari, Mehrbakhsh Nilashi,
  and Mojtaba Alizadeh.
\newblock The application of internet of things in healthcare: a systematic
  literature review and classification.
\newblock {\em Universal Access in the Information Society}, 18(4):837--869,
  2019.

\bibitem{92}
N.~{Hema Rajini}.
\newblock A comprehensive survey on internet of things based healthcare
  services and its applications.
\newblock In {\em 2019 3rd International Conference on Computing Methodologies
  and Communication (ICCMC)}, pages 483--488, 2019.

\bibitem{91}
H.~{Habibzadeh}, K.~{Dinesh}, O.~{Rajabi Shishvan}, A.~{Boggio-Dandry},
  G.~{Sharma}, and T.~{Soyata}.
\newblock A survey of healthcare internet of things (hiot): A clinical
  perspective.
\newblock {\em IEEE Internet of Things Journal}, 7(1):53--71, 2020.

\bibitem{94}
Muhammet Usak, Milan Kubiatko, Muhammad~Salman Shabbir, Olesya
  Viktorovna~Dudnik, Kittisak Jermsittiparsert, and Lila Rajabion.
\newblock Health care service delivery based on the internet of things: A
  systematic and comprehensive study.
\newblock {\em International Journal of Communication Systems}, 33(2):e4179,
  2020.
\newblock e4179 IJCS-19-0281.R1.

\bibitem{10}
Mrinai~M. Dhanvijay and Shailaja~C. Patil.
\newblock Internet of things: A survey of enabling technologies in healthcare
  and its applications.
\newblock {\em Computer Networks}, 153:113--131, 2019.

\bibitem{99}
Yazdan~Ahmad Qadri, Ali Nauman, Yousaf~Bin Zikria, Athanasios~V. Vasilakos, and
  Sung~Won Kim.
\newblock The future of healthcare internet of things: A survey of emerging
  technologies.
\newblock {\em IEEE Communications Surveys Tutorials}, 22(2):1121--1167, 2020.

\bibitem{84}
M.~l.~{Sahu}, M.~{Atulkar}, and M.~K. {Ahirwal}.
\newblock Comprehensive investigation on iot based smart healthcare system.
\newblock In {\em 2020 First International Conference on Power, Control and
  Computing Technologies (ICPC2T)}, pages 325--330, 2020.

\bibitem{65}
A.~{Kumar}, R.~{Krishnamurthi}, A.~{Nayyar}, K.~{Sharma}, V.~{Grover}, and
  E.~{Hossain}.
\newblock A novel smart healthcare design, simulation, and implementation using
  healthcare 4.0 processes.
\newblock {\em IEEE Access}, 8:118433--118471, 2020.

\bibitem{66}
B.~{Mohanta}, P.~{Das}, and S.~{Patnaik}.
\newblock Healthcare 5.0: A paradigm shift in digital healthcare system using
  artificial intelligence, iot and 5g communication.
\newblock In {\em 2019 International Conference on Applied Machine Learning
  (ICAML)}, pages 191--196, 2019.

\bibitem{85}
T.~{Akca}, O.~K. {Sahingoz}, E.~{Kocyigit}, and M.~{Tozal}.
\newblock Intelligent ambulance management system in smart cities.
\newblock In {\em 2020 International Conference on Electrical Engineering
  (ICEE)}, pages 1--7, 2020.

\bibitem{11}
Bahar Farahani, Farshad Firouzi, Victor Chang, Mustafa Badaroglu, Nicholas
  Constant, and Kunal Mankodiya.
\newblock Towards fog-driven iot ehealth: Promises and challenges of iot in
  medicine and healthcare.
\newblock {\em Future Generation Computer Systems}, 78:659--676, 2018.

\bibitem{62}
S.~{Chatterjee}, S.~{Chatterjee}, S.~{Choudhury}, S.~{Basak}, S.~{Dey},
  S.~{Sain}, K.~S. {Ghosal}, N.~{Dalmia}, and S.~{Sircar}.
\newblock Internet of things and body area network-an integrated future.
\newblock In {\em 2017 IEEE 8th Annual Ubiquitous Computing, Electronics and
  Mobile Communication Conference (UEMCON)}, pages 396--400, 2017.

\bibitem{79}
F.~{Sallabi}, F.~{Naeem}, M.~{Awad}, and K.~{Shuaib}.
\newblock Managing iot-based smart healthcare systems traffic with software
  defined networks.
\newblock In {\em 2018 International Symposium on Networks, Computers and
  Communications (ISNCC)}, pages 1--6, 2018.

\bibitem{46}
T.~{Zhang}, A.~H. {Sodhro}, Z.~{Luo}, N.~{Zahid}, M.~W. {Nawaz},
  S.~{Pirbhulal}, and M.~{Muzammal}.
\newblock A joint deep learning and internet of medical things driven framework
  for elderly patients.
\newblock {\em IEEE Access}, 8:75822--75832, 2020.

\bibitem{67}
H.~{Harb}, A.~{Mansour}, A.~{Nasser}, E.~M. {Cruz}, and I.~{de la Torre Díez}.
\newblock A sensor-based data analytics for patient monitoring in connected
  healthcare applications.
\newblock {\em IEEE Sensors Journal}, 21(2):974--984, 2021.

\bibitem{76}
G.~{Tsang}, X.~{Xie}, and S.~M. {Zhou}.
\newblock Harnessing the power of machine learning in dementia informatics
  research: Issues, opportunities, and challenges.
\newblock {\em IEEE Reviews in Biomedical Engineering}, 13:113--129, 2020.

\bibitem{81}
W.~{Qi}, H.~{Su}, and A.~{Aliverti}.
\newblock A smartphone-based adaptive recognition and real-time monitoring
  system for human activities.
\newblock {\em IEEE Transactions on Human-Machine Systems}, 50(5):414--423,
  2020.

\bibitem{71}
M.~S. {Hossain} and G.~{Muhammad}.
\newblock Deep learning based pathology detection for smart connected
  healthcare.
\newblock {\em IEEE Network}, 34(6):120--125, 2020.

\bibitem{78}
P.~{Dong}, Z.~{Ning}, M.~S. {Obaidat}, X.~{Jiang}, Y.~{Guo}, X.~{Hu}, B.~{Hu},
  and B.~{Sadoun}.
\newblock Edge computing based healthcare systems: Enabling decentralized
  health monitoring in internet of medical things.
\newblock {\em IEEE Network}, 34(5):254--261, 2020.

\bibitem{48}
Alcardo~Alex Barakabitze, A.~Ahmad, Rashid Mijumbi, and A.~Hines.
\newblock 5g network slicing using sdn and nfv- a survey of taxonomy,
  architectures and future challenges.
\newblock {\em Comput. Networks}, 167, 2020.

\bibitem{58}
E.~{Gures}, I.~{Shayea}, A.~{Alhammadi}, M.~{Ergen}, and H.~{Mohamad}.
\newblock A comprehensive survey on mobility management in 5g heterogeneous
  networks: Architectures, challenges and solutions.
\newblock {\em IEEE Access}, 8:195883--195913, 2020.

\bibitem{77}
J.~{Li}, J.~{Cai}, F.~{Khan}, A.~U. {Rehman}, V.~{Balasubramaniam}, J.~{Sun},
  and P.~{Venu}.
\newblock A secured framework for sdn-based edge computing in iot-enabled
  healthcare system.
\newblock {\em IEEE Access}, 8:135479--135490, 2020.

\bibitem{70}
E.~M. {Abou-Nassar}, A.~M. {Iliyasu}, P.~M. {El-Kafrawy}, O.~{Song}, A.~K.
  {Bashir}, and A.~A.~A. {El-Latif}.
\newblock Ditrust chain: Towards blockchain-based trust models for sustainable
  healthcare iot systems.
\newblock {\em IEEE Access}, 8:111223--111238, 2020.

\bibitem{74}
J.~{Xu}, K.~{Xue}, S.~{Li}, H.~{Tian}, J.~{Hong}, P.~{Hong}, and N.~{Yu}.
\newblock Healthchain: A blockchain-based privacy preserving scheme for
  large-scale health data.
\newblock {\em IEEE Internet of Things Journal}, 6(5):8770--8781, 2019.

\bibitem{9}
V.~P. Rachim and W.~Chung.
\newblock Wearable noncontact armband for mobile ecg monitoring system.
\newblock {\em IEEE Transactions on Biomedical Circuits and Systems},
  10(6):1112--1118, 2016.

\bibitem{82}
H.~T. {Yew}, M.~F. {Ng}, S.~Z. {Ping}, S.~K. {Chung}, A.~{Chekima}, and J.~A.
  {Dargham}.
\newblock Iot based real-time remote patient monitoring system.
\newblock In {\em 2020 16th IEEE International Colloquium on Signal Processing
  Its Applications (CSPA)}, pages 176--179, 2020.

\bibitem{87}
D.~{Zhong}, Z.~{Yian}, W.~{Lanqing}, D.~{Junhua}, and H.~{Jiaxuan}.
\newblock Continuous blood pressure measurement platform: A wearable system
  based on multidimensional perception data.
\newblock {\em IEEE Access}, 8:10147--10158, 2020.

\bibitem{86}
I.~{Bisio}, C.~{Garibotto}, F.~{Lavagetto}, and A.~{Sciarrone}.
\newblock When ehealth meets iot: A smart wireless system for post-stroke home
  rehabilitation.
\newblock {\em IEEE Wireless Communications}, 26(6):24--29, 2019.

\bibitem{88}
C.~{Sandeepa}, C.~{Moremada}, N.~{Dissanayaka}, T.~{Gamage}, and M.~{Liyanage}.
\newblock An emergency situation detection system for ambient assisted living.
\newblock In {\em 2020 IEEE International Conference on Communications
  Workshops (ICC Workshops)}, pages 1--6, 2020.

\bibitem{ahmadi2019application}
Hossein Ahmadi, Goli Arji, Leila Shahmoradi, Reza Safdari, Mehrbakhsh Nilashi,
  and Mojtaba Alizadeh.
\newblock The application of internet of things in healthcare: a systematic
  literature review and classification.
\newblock {\em Universal Access in the Information Society}, pages 1--33, 2019.

\bibitem{25}
S.~M.~R. {Islam}, D.~{Kwak}, M.~H. {Kabir}, M.~{Hossain}, and K.~{Kwak}.
\newblock The internet of things for health care: A comprehensive survey.
\newblock {\em IEEE Access}, 3:678--708, 2015.

\bibitem{89}
Victoria Ahmadi, Sophia Benjelloun, Michel El~Kik, Tanvi Sharma, Huihui Chi,
  and Wei Zhou.
\newblock Drug governance: Iot-based blockchain implementation in the
  pharmaceutical supply chain.
\newblock pages 1--8, 02 2020.

\bibitem{101}
Mohammad Jabirullah, Rakesh Ranjan, Mirza~Nemath Ali~Baig, and Anish
  Kumar~Vishwakarma.
\newblock Development of e-health monitoring system for remote rural community
  of india.
\newblock In {\em 2020 7th International Conference on Signal Processing and
  Integrated Networks (SPIN)}, pages 767--771, 2020.

\bibitem{100}
Prabha Sundaravadivel, Issac Lee, Saraju Mohanty, Elias Kougianos, and Laavanya
  Rachakonda.
\newblock Rm-iot: An iot based rapid medical response plan for smart cities.
\newblock In {\em 2019 IEEE International Symposium on Smart Electronic Systems
  (iSES) (Formerly iNiS)}, pages 241--246, 2019.

\bibitem{34}
T.~{Shaown}, I.~{Hasan}, M.~M.~R. {Mim}, and M.~S. {Hossain}.
\newblock Iot-based portable ecg monitoring system for smart healthcare.
\newblock In {\em 2019 1st International Conference on Advances in Science,
  Engineering and Robotics Technology (ICASERT)}, pages 1--5, 2019.

\bibitem{35}
Pramila Arulanthu and Eswaran Perumal.
\newblock An intelligent iot with cloud centric medical decision support system
  for chronic kidney disease prediction.
\newblock {\em International Journal of Imaging Systems and Technology}, 03
  2020.

\bibitem{36}
Priyan~Malarvizhi Kumar and Usha {Devi Gandhi}.
\newblock A novel three-tier internet of things architecture with machine
  learning algorithm for early detection of heart diseases.
\newblock {\em Computers \& Electrical Engineering}, 65:222 -- 235, 2018.

\bibitem{72}
J.~{Baek} and K.~{Chung}.
\newblock Context deep neural network model for predicting depression risk
  using multiple regression.
\newblock {\em IEEE Access}, 8:18171--18181, 2020.

\bibitem{37}
T.~{Tekeste}, H.~{Saleh}, B.~{Mohammad}, and M.~{Ismail}.
\newblock Ultra-low power qrs detection and ecg compression architecture for
  iot healthcare devices.
\newblock {\em IEEE Transactions on Circuits and Systems I: Regular Papers},
  66(2):669--679, 2019.

\bibitem{42}
U.~{Satija}, B.~{Ramkumar}, and M.~{Sabarimalai Manikandan}.
\newblock Real-time signal quality-aware ecg telemetry system for iot-based
  health care monitoring.
\newblock {\em IEEE Internet of Things Journal}, 4(3):815--823, June 2017.

\bibitem{83}
G.~{Yang}, M.~A. {Jan}, V.~G. {Menon}, P.~G. {Shynu}, M.~M. {Aimal}, and M.~D.
  {Alshehri}.
\newblock A centralized cluster-based hierarchical approach for green
  communication in a smart healthcare system.
\newblock {\em IEEE Access}, 8:101464--101475, 2020.

\bibitem{17}
A.~B. {Pawar} and S.~{Ghumbre}.
\newblock A survey on iot applications, security challenges and counter
  measures.
\newblock In {\em 2016 International Conference on Computing, Analytics and
  Security Trends (CAST)}, pages 294--299, 2016.

\bibitem{41}
N.~{Karimian}, M.~{Tehranipoor}, D.~{Woodard}, and D.~{Forte}.
\newblock Unlock your heart: Next generation biometric in resourceconstrained
  healthcare systems and iot.
\newblock {\em IEEE Access}, 7:49135--49149, 2019.

\bibitem{43}
M.~{Min}, X.~{Wan}, L.~{Xiao}, Y.~{Chen}, M.~{Xia}, D.~{Wu}, and H.~{Dai}.
\newblock Learning-based privacy-aware offloading for healthcare iot with
  energy harvesting.
\newblock {\em IEEE Internet of Things Journal}, 6(3):4307--4316, 2019.

\bibitem{45}
Yang Yang, Xianghan Zheng, Wenzhong Guo, Ximeng Liu, and Victor Chang.
\newblock Privacy-preserving smart iot-based healthcare big data storage and
  self-adaptive access control system.
\newblock {\em Information Sciences}, 479:567 -- 592, 2019.

\bibitem{50}
Geetanjali Rathee, Ashutosh Sharma, Hemraj Saini, Rajiv Kumar, and Razi Iqbal.
\newblock A hybrid framework for multimedia data processing in iot-healthcare
  using blockchain technology.
\newblock {\em Multimedia Tools and Applications}, 06 2019.

\bibitem{38}
C.~{Yi} and J.~{Cai}.
\newblock A truthful mechanism for scheduling delay-constrained wireless
  transmissions in iot-based healthcare networks.
\newblock {\em IEEE Transactions on Wireless Communications}, 18(2):912--925,
  2019.

\bibitem{44}
K.~{Wang}, Y.~{Shao}, L.~{Xie}, J.~{Wu}, and S.~{Guo}.
\newblock Adaptive and fault-tolerant data processing in healthcare iot based
  on fog computing.
\newblock {\em IEEE Transactions on Network Science and Engineering},
  7(1):263--273, 2020.

\bibitem{39}
R.~K. {Pathinarupothi}, P.~{Durga}, and E.~S. {Rangan}.
\newblock Iot-based smart edge for global health: Remote monitoring with
  severity detection and alerts transmission.
\newblock {\em IEEE Internet of Things Journal}, 6(2):2449--2462, 2019.

\bibitem{57}
Suneeta~S. Raykar and Vinayak~N. Shet.
\newblock Design of healthcare system using iot enabled application.
\newblock {\em Materials Today: Proceedings}, 23:62 -- 67, 2020.
\newblock Advanced Materials for Clean Energy and Health Applications (AMCEHA)
  , University of Jaffna, Jafna, Sri Lanka, 6-8 February, 2019.

\bibitem{61}
Y.~A. {Qadri}, A.~{Nauman}, Y.~B. {Zikria}, A.~V. {Vasilakos}, and S.~W. {Kim}.
\newblock The future of healthcare internet of things: A survey of emerging
  technologies.
\newblock {\em IEEE Communications Surveys Tutorials}, 22(2):1121--1167, 2020.

\bibitem{56}
Farhan Ullah, Muhammad~Asif Habib, Muhammad Farhan, Shehzad Khalid, Mehr~Yahya
  Durrani, and Sohail Jabbar.
\newblock Semantic interoperability for big-data in heterogeneous iot
  infrastructure for healthcare.
\newblock {\em Sustainable Cities and Society}, 34:90 -- 96, 2017.

\bibitem{63}
P.~{Pace}, G.~{Aloi}, G.~{Caliciuri}, R.~{Gravina}, C.~{Savaglio},
  G.~{Fortino}, G.~{Ibanez-Sanchez}, A.~{Fides-Valero}, J.~{Bayo-Monton},
  M.~{Uberti}, M.~{Corona}, L.~{Bernini}, M.~{Gulino}, A.~{Costa}, I.~{De
  Luca}, and M.~{Mortara}.
\newblock Inter-health: An interoperable iot solution for active and assisted
  living healthcare services.
\newblock In {\em 2019 IEEE 5th World Forum on Internet of Things (WF-IoT)},
  pages 81--86, 2019.

\bibitem{31}
Manikandan Rajasekaran, Abdulsalam Yassine, M.~Shamim Hossain, Mohammed~F.
  Alhamid, and Mohsen Guizani.
\newblock Autonomous monitoring in healthcare environment: Reward-based energy
  charging mechanism for iomt wireless sensing nodes.
\newblock {\em Future Generation Computer Systems}, 98:565 -- 576, 2019.

\bibitem{33}
Hodjat Hamidi.
\newblock An approach to develop the smart health using internet of things and
  authentication based on biometric technology.
\newblock {\em Future Generation Computer Systems}, 91:434 -- 449, 2019.

\bibitem{32}
Azadeh Zamanifar and Eslam Nazemi.
\newblock An approach for predicting health status in iot health care.
\newblock {\em Journal of Network and Computer Applications}, 134:100--113, 03
  2019.

\bibitem{40}
W.~{Manatarinat}, S.~{Poomrittigul}, and P.~{Tantatsanawong}.
\newblock Narrowband-internet of things (nb-iot) system for elderly healthcare
  services.
\newblock In {\em 2019 5th International Conference on Engineering, Applied
  Sciences and Technology (ICEAST)}, pages 1--4, 2019.

\bibitem{60}
V.~K. {Daliya} and T.~K. {Ramesh}.
\newblock Data interoperability enhancement of electronic health record data
  using a hybrid model.
\newblock In {\em 2019 International Conference on Smart Systems and Inventive
  Technology (ICSSIT)}, pages 318--322, 2019.

\bibitem{13}
J.~M.~Camara Brito.
\newblock Trends in wireless communications towards 5g networks — the
  influence of e-health and iot applications.
\newblock In {\em 2016 International Multidisciplinary Conference on Computer
  and Energy Science (SpliTech)}, pages 1--7.

\bibitem{14}
P.~A.~H. Williams and V.~McCauley.
\newblock Always connected: The security challenges of the healthcare internet
  of things.
\newblock In {\em 2016 IEEE 3rd World Forum on Internet of Things (WF-IoT)},
  pages 30--35.

\bibitem{64}
X.~{Zhou}, W.~{Liang}, K.~I. {Wang}, H.~{Wang}, L.~T. {Yang}, and Q.~{Jin}.
\newblock Deep-learning-enhanced human activity recognition for internet of
  healthcare things.
\newblock {\em IEEE Internet of Things Journal}, 7(7):6429--6438, 2020.

\bibitem{73}
Joshua Raj, S.~Shobana, Irina Pustokhina, Denis Pustokhin, Deepak Gupta, and
  Dr~Shankar.
\newblock Optimal feature selection-based medical image classification using
  deep learning model in internet of medical things.
\newblock {\em IEEE Access}, PP:1--1, 03 2020.

\bibitem{68}
A.~{Sciarrone}, I.~{Bisio}, C.~{Garibotto}, F.~{Lavagetto}, G.~H. {Staude}, and
  A.~{Knopp}.
\newblock Leveraging iot wearable technology towards early diagnosis of
  neurological diseases.
\newblock {\em IEEE Journal on Selected Areas in Communications},
  39(2):582--592, 2021.

\bibitem{75}
M.~S. {Hossain}, G.~{Muhammad}, and N.~{Guizani}.
\newblock Explainable ai and mass surveillance system-based healthcare
  framework to combat covid-i9 like pandemics.
\newblock {\em IEEE Network}, 34(4):126--132, 2020.

\end{thebibliography}

\vfill

\end{document}